УДК 681.3

Крамов А. А., студент,
Баужа О. С., к.ф.-м.н., асистент

A. A. Kramov, student,
O. S. Bauzha, PhD.

## Розробка програмно-апаратного комплексу віддаленого моніторингу пульсу людини

## Development of the complex system for the remote monitoring of the human heart rate

Київський національний університет імені Тараса Шевченка, 83000, м. Київ, пр-т. Глушкова 4г,
e-mail: artemkramov@gmail.com
e-mail: asb@mail.univ.kiev.ua

Taras Shevchenko National University of Kyiv, 83000, Kyiv, Glushkova st., 4g,
e-mail: artemkramov@gmail.com
e-mail: asb@mail.univ.kiev.ua

*В роботі пропонується метод створення автоматизованого комплексу віддаленого спостереження та аналізу пульсу людини. Результатом роботи є розробка програмно-апаратного комплексу моніторингу пульсу людини в режимі реального часу. Комплекс може здійснювати візуальну демонстрацію процесу серцебиття з використанням веб-інтерфейсу та оповіщати компетентних осіб у випадку відхилення значення пульсу від допустимих меж.*
*Ключові слова: пульс, ATmega8, Arduino, Bluetooth.*

*An implementation of the remote pulse monitoring system which allows observing of the patient's pulse in the real-time mode via browser is offered in this work. The result of the work is the development of the complex system, which contains the hardware components for the pulse measurement and the software component for the data processing and visualization in a web-interface. The web-interface provides the heart rate visualization in the real-time mode and informs the appropriate person in case of deviation from pulse limits. The monitoring system can detect two disease types: tachycardia and bradycardia. A pulse sensor detects the heart beat moment and functions like a plethysmograph. The microcontroller ATmega8 is used to read data from the sensor, to analyze information and pass it to the next hardware block. Arduino Uno and Ethernet module ENC28J60 are used to transform the information about the heart beat event to the web-interface. Ethernet module ENC28J60 is connected to Arduino Uno using SPI interface. The pair of Bluetooth modules HC-05 is used to connect ATmega8 and Arduino Uno between each other. The module HC-05 is connected to both microcontrollers using UART interface. The Websocket protocol is used to implement the real-time data demonstration in the web-interface. The web-interface is adapted to mobile devices therefore it can be viewed from smartphones and tablets. The complex can be used both by the qualified specialist for the remote monitoring of the patient's state and as personal prophylactic.*

Статтю представив д.т.н. Погорілий С.Д.

### Вступ

Серцево-судинні захворювання – найпоширеніша медично-біологічна і соціальна проблема. Даний тип захворювання є найтиповішою причиною смертності в усьому світі, і в Україні в тому числі. Станом на 2016 рік Україна знаходиться серед країн з найбільшою кількістю серцево-судинних хвороб в Європі [1]. За статистичними даними [1] відносна частка людей в Україні, що страждають на серцево-судинні захворювання, становить близько 19%. Лікування, запобігання, профілактика серцево-судинних хвороб є актуальними проблемами в Україні та потребують якнайшвидшого вирішення.

Для запобігання ускладнень та профілактики серцево-судинних хвороб необхідне постійне спостереження за серцевим ритмом людини та оперативна реакція на його відхилення від норми. Аналіз пульсу людини дозволяє відслідкувати ряд хвороб: тахікардію (пульс вище норми), брадикардію (пульс нижче норми), аритмію. Багато людей похилого віку мають проблеми з порушенням пам'яті чи мислення, тому у випадку погіршення стану здоров'я не в змозі надати собі допомогу, чи, інколи, навіть звернутися за допомогою. Найгірша ситуація, коли хвора людина одинока. Тому реалізація моніторингу пульсу людини у режимі реального часу є актуальною проблемою.

Існують програмно-апаратні системи, що дозволяються виконувати дистанційне спостереження пульсу, проте кожна з них має власні недоліки: необхідність в наявності мобільного пристрою [2] чи NFC-сумісного (NFC – Near Field Communication) пристрою [3], складність трансляції даних в мережу. Варто відзначити високу вартість таких пристроїв, а також відсутність їхніх аналогів вітчизняного виробництва.

В роботі пропонується створення програмно-апаратного комплексу для можливості віддаленого моніторингу та аналізу пульсу людини в режимі реального часу.

Під час розробки комплексу до апаратної частини були висунуті наступні вимоги:

– наявність пристрою отримання даних про пульс (періодичність серцевих скорочень);
– можливість передачі даних на веб-сервер.

На програмну частину покладено наступний функціонал:

– зчитування даних з апаратної частини та вивід даних у веб-інтерфейсі в режимі реального часу;
– аналіз пульсу та оповіщення компетентних осіб у випадку відхилення від допустимих меж.

**Основні принципи роботи комплексу**

В цілому комплекс можна розділити на 3 функціональні блоки, з'єднані послідовно:

– апаратний блок зчитування даних про пульс людини;
– апаратний блок передачі даних до веб-служби;
– програмний блок отримання, аналізу та візуального відтворення пульсу людини.

На рис.1 наведена загальна блок-схема програмно-апаратного комплексу та відповідні функціональні блоки.

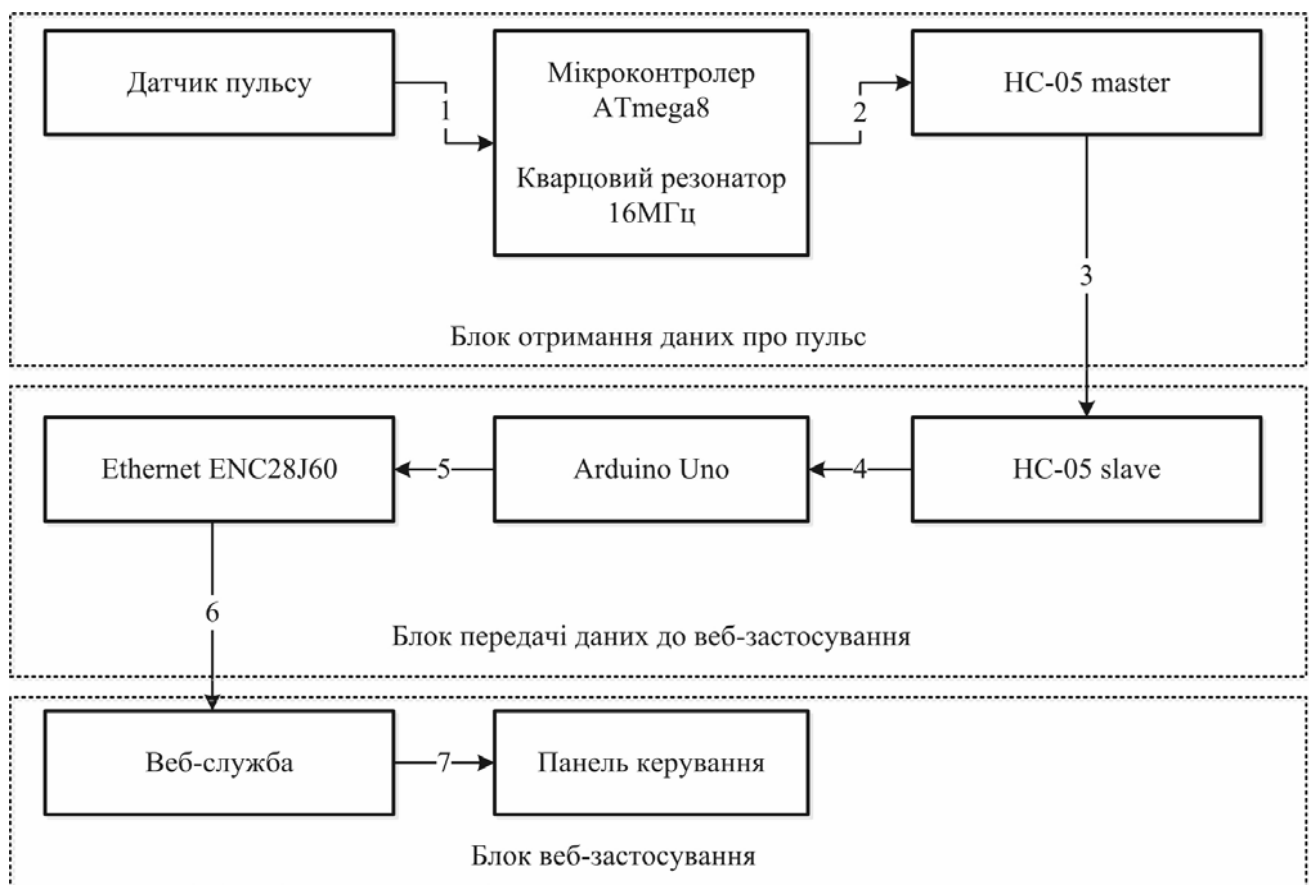

Рис. 1 Блок-схема програмно-апаратного комплексу

Перший блок забезпечує отримання інформації про пульс. Елементи даного блоку разом складають єдиний елемент, який кріпиться на руку людині. Схема першого блоку зображена на рис. 2. Другий блок відповідальний за передачу інформації в мережу, що здійснюється за допомогою Ethernet-модуля. Схема другого блоку зображена на рис. 3. Третій блок здійснює обробку отриманих даних та їхню візуалізацію. Візуалізація здійснена завдяки веб-інтерфейсу, що надає можливість перегляду поточного пульсу пацієнта, діаграму пульсу за визначений період, статистику захворювань.

На початковому етапі дані зчитуються з датчику пульсу [4] мікроконтролером ATmega8 [5]. Принцип роботи датчика оснований на методі фотоплетизмографії. Плетизмографія – вимірювання пульсу на основі реєстрації зміни об'єму в органі, наповненому кров'ю (як правило – в кровоносних судинах). Під фотоплетизмографією розуміється реєстрація зміни наповненості частини тіла кров'ю за допомогою оптичних методів. Вимірювання оптичної густини відбувається за допомогою фотоприймача та джерела випромінювання, зокрема світлодіоду. Вимірювання, як правило, проводиться з пальця руки або мочки вуха.

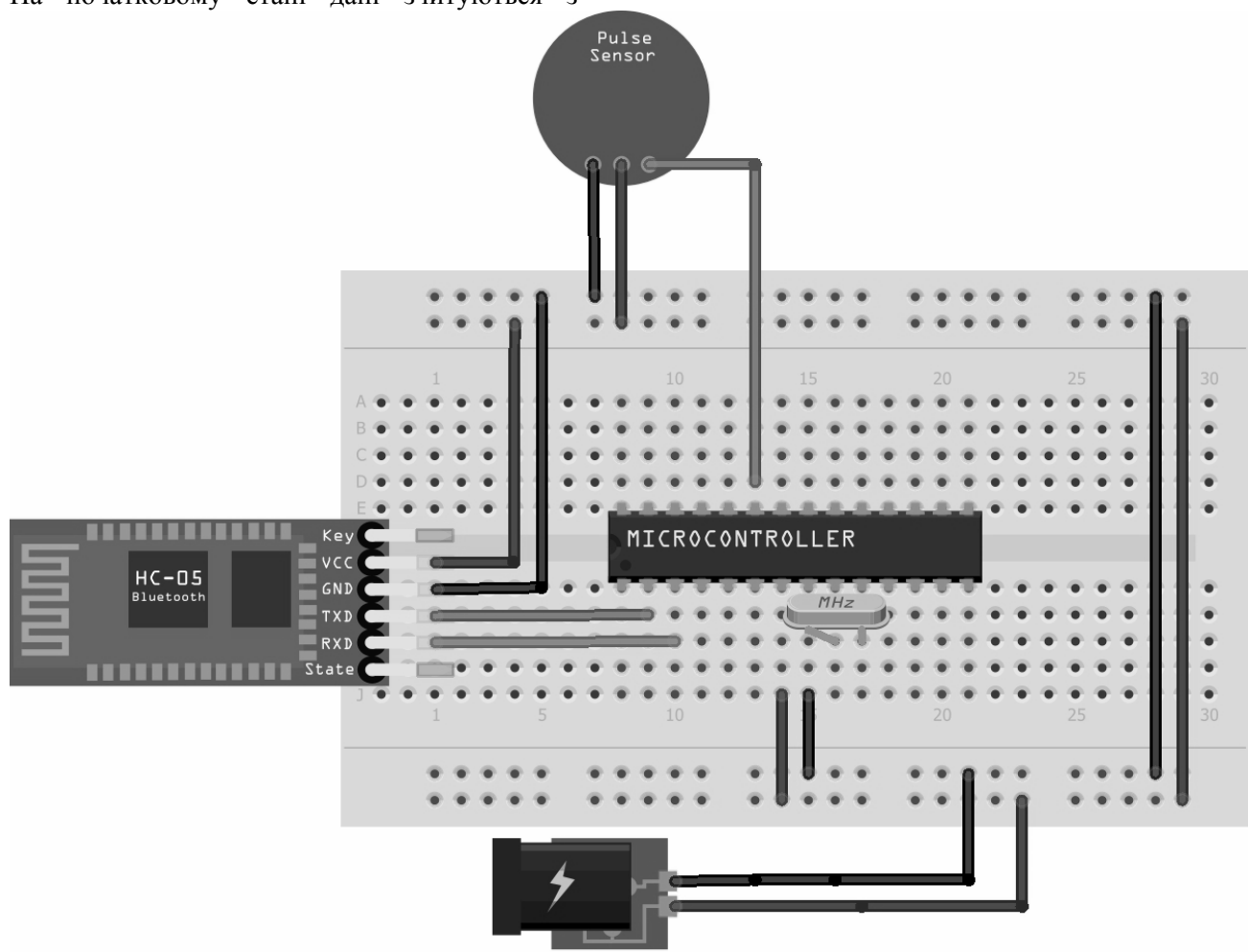

Рис. 2 Схема блоку реєстрації пульсу людини

По інтенсивності світіння джерела можна зробити висновок про поточний об'єм крові в досліджуваній частині тіла.

Після реєстрації моменту удару серця дані необхідно передати до наступного блоку комплексу, забезпечуючи достатню відстань передачі. Для цього використовується пара Bluetooth-пристроїв серії HC-05 [6]. Один з пристроїв працює в режимі передавача (див. рис. 2), інший в режимі приймача (див. рис. 3). Головними перевагами інтерфейсу Bluetooth є стійкість до широкосмугових завад (сукупність пристроїв, що знаходяться в одному місці, можуть одночасно здійснювати обмін даними між собою без впливу на сусідні канали зв'язку) та простота реалізації. Для передачі даних між пристроями необхідно встановити між ними захищене з'єднання, що реалізується за допомогою АТ-команд [7]. АТ-команди дозволяють встановити роль пристрою (передавач чи приймач), пароль з'єднання та зробити пристрої спряженими. Спряжені пристрої встановлюють захищений канал зв'язку між собою, що робить з'єднання стійким до зовнішніх завад.

Обмін даними з пристроєм HC-05 здійснюється за допомогою інтерфейсу UART (Universal Asynchronous Receiver/Transmitter) [8]. UART є популярним інтерфейсом асинхронної передачі даних, що з'єднує компоненти комп'ютерів та периферійних пристроїв. Реалізація UART наявна в багатьох серія мікроконтролерів, включаючи ATmega8. Швидкість передачі даних між мікроконтролером та Bluetooth-пристроєм складає значення 38600 бод.

Для того, щоб синхронізація передачі даних відбулася успішно, необхідно, щоб тактова частота джерела передачі сигналу та приймача була однаковою. Тактова частота процесора плати для налагодження Arduino Uno [9] складає значення 16МГц, мікроконтролера ATmega8 – 8МГц. Тому як джерело тактової частоти мікроконтролера ATmega8 використовується кварцовий резонатор (див. рис.2) з робочою частотою 16МГц.

Для обробки даних, отриманих з Bluetooth приймача, використовується плата Arduino Uno. Arduino Uno [9] (див. рис.3) – апаратно-обчислювальна, відкрита платформа, призначена для швидкого конструювання різнотипних електронних пристроїв. Платформа є доволі популярною у світі завдяки зручності у використанні та простоті мови програмування. Для програмування використовується модифікована версія C++, відома як Wiring. Як середовище програмування можна використовувати водночас вільно поширюване середовище Arduino IDE, а також будь-який C/C++ інструментарій (наприклад, Atmel Studio).

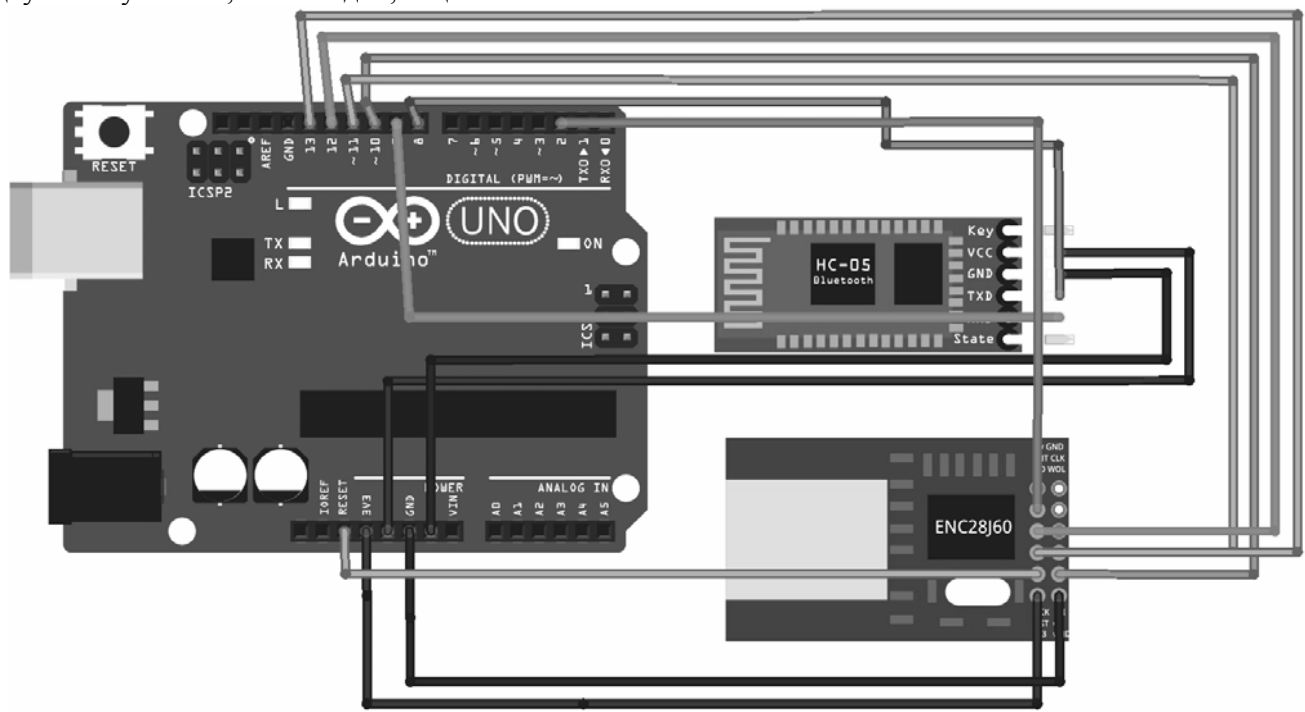

Рис. 3 Схема блоку передачі даних в мережу

Для розробки програмного забезпечення можна використовувати операційні системи Windows, Linux, MacOS X. Плата має вбудований програматор, в який записано завантажувач (bootloader), тому немає потреби у використанні зовнішнього програматора для зміни програмного забезпечення. До складу плати входять наступні компоненти: 14 цифрових входів/виходів (з них 6 можуть використовуватися як виходи для широтно-імпульсної модуляції), 6 аналогових входів, кварцовий резонатор на 16МГц, роз'єм USB, роз'єм живлення, роз'єм для внутрішньо-схемного програмування (ICSP) і кнопка скидання. Для керування роботою компонентів плати використовується мікроконтролер ATmega328 [10]. Як перетворювач інтерфейсів USB-UART застосовується мікроконтролер ATmega16U2. Для початку роботи з пристроєм необхідно подати напругу від AC/DC-адаптера чи батарейки, або підключити його до комп'ютера через USB-з'єднання. Схемотехнічна реалізація платформи Arduino Uno поширена у відкритому доступі.

Передача даних з Arduino Uno в мережу відбувається за допомогою Ethernet-модуля ENC28J60. Модуль може працювати водночас в режимі HTTP-сервера, а також як HTTP-клієнт. Зв'язок з мікроконтролером відбувається через інтерфейс SPI (Serial Peripheral Interface). Робоча напруга пристрою – 3.3В. На платі змонтована розетка RJ-45 для кабелю Ethernet. Розетка містить трансформатор, 2 світлодіоди, що є індикаторами обміну даними, кварцовий резонатор частоти 25МГц.

Дані, представлені у форматі JSON (JavaScript Object Notation), апаратним блоком передаються на веб-сервіс, який обробляє, зберігає та відправляє дані у веб-інтерфейс для подальшого візуального представлення. Для відображення діаграми пульсу в режимі реального часу необхідне постійне з'єднання веб-сервісу з веб-інтерфейсом. Як механізм перманентного з'єднання використовуються веб-сокети і протокол передачі даних WebSocket. WebSocket — це протокол транспортного рівня, призначений для обміну інформацією між браузером та веб-сервером в режимі реального часу. Він забезпечує двонаправлений повнодуплексний канал зв'язку через один TCP-сокет. Протокол WebSocket спроектовано для втілення у веб-браузерах та веб-серверах, але може також використовуватись будь-якою клієнт-серверною програмою. Прикладний програмний інтерфейс WebSocket був стандартизований міжнародним співтовариством W3C, крім того протокол WebSocket стандартизований організацією IETF як документ RFC 6455. Зв'язок здійснюється через TCP порт 80, а отже, його можна застосовувати в тих середовищах, де інші мережеві протоколи заблоковано.

Розроблений пристрій є простим у повторенні, адже для реалізації апаратної частини використовувалися стандартні та широко вживані серед фахівців компоненти. Вартість одного приладу при малосерійному виготовленні складає близько 20$.

## Висновки

Велика поширеність серцево-судинних хвороб в Україні, де статистика серцевих захворювань є найгіршою в Європі, підтверджує той факт, що профілактика та запобігання серцево-судинних хвороб є важливою і актуальною проблемою. Розробка програмно-апаратного комплексу, здатного віддалено здійснювати постійний моніторинг пульсу людини та передбачати можливі загрози, дозволить водночас запобігти можливим серцевим захворюванням, а також швидко зреагувати на погіршення стану людей, не здатних надати собі допомогу самотужки.

Серед різних проаналізованих методів вимірювання пульсу перевагу надано методу фотоплетизмографії. Відповідний датчик пульсу має достатню точність та є доступним за вартістю. В ролі обчислювального пристрою для отримання даних з датчика обрано мікроконтролер ATmega8, характерний невеликими розмірами та низьким енергоспоживанням. Для передачі даних в мережу використана популярна плата для налагодження Arduino Uno та Ethernet-модуль ENC28J60.

Продемонстровано, що для бездротового зв'язку мікроконтролерів зручно використовувати Bluetooth пристрої серії HC-05, що працюють як передавач та приймач інформації. Пара таких спряжених пристроїв має достатньо великий запас швидкодії, низьке енергоспоживання та є простою у використанні.

Оскільки моніторинг пульсу повинен відбуватися в режимі реального часу, необхідне постійне і надійне з'єднання пристрою з веб-інтерфейсом. Для реалізації перманентного з'єднання зручно використовувати протокол WebSocket, призначений для обміну повідомленнями між браузером і веб-сервером в режимі реального часу.

Розроблений програмно-апаратний комплекс може використовуватися компетентним фахівцем для віддаленого слідкування за станом пацієнта, а також як персональний профілактичний засіб.